# The Anatomy of a Grid portal

**Daniele Licari**[1,2], **Federico Calzolari**[1,3]
[1] Scuola Normale Superiore, Pisa, Italy
[2] University of Pisa, Computer Science Department, Pisa, Italy
[3] Italian National Institute for Nuclear Physics INFN, Pisa, Italy

E-mail: daniele.licari@sns.it, federico.calzolari@sns.it

**Abstract.** In this paper we introduce a new way to deal with Grid portals referring to our implementation. L-GRID is a light portal to access the EGEE/EGI Grid infrastructure via Web, allowing users to submit their jobs from a common Web browser in a few minutes, without any knowledge about the Grid infrastructure. It provides the control over the complete lifecycle of a Grid Job, from its submission and status monitoring, to the output retrieval. The system, implemented as client-server architecture, is based on the Globus Grid middleware. The client side application is based on a java applet; the server relies on a Globus User Interface. There is no need of user registration on the server side, and the user needs only his own X.509 personal certificate. The system is user-friendly, secure (it uses SSL protocol, mechanism for dynamic delegation and identity creation in public key infrastructures), highly customizable, open source, and easy to install. The X.509 personal certificate does not get out from the local machine. It allows to reduce the time spent for the job submission, granting at the same time a higher efficiency and a better security level in proxy delegation and management.

## 1. Introduction
While most of the research, both in science and humanities, requires a growing computational power and data storage, the access to the Grid computing resources is nowadays too complex for a non-practiced researcher. Grid is today a very powerful tool for a few people, usually involved in big projects in physics, computational chemistry, biomedicine [1,2].
Use simplification has today become a common practice in the access and utilization of Cloud, Grid, and Datacenter resources [3,4]. Our idea is to give users a very simple tool to access the EGEE/EGI Grid resources, without requiring users to be expert in computer science or distributed computing architecture. We deployed a Web portal, accessible via Web browser, in order to provide users with a simple interface to the Grid world. L-GRID is a light portal to access Grid infrastructure via Web browser, allowing users to submit their jobs in a few minutes, without any knowledge about the Grid infrastructure.
L-GRID provides the typical operations involved in a Grid environment: certificate conversion, job submission, job status monitoring, and output retrieval. It provides also a JDL editor. The system is user-friendly, secure - it uses SSL protocol, mechanism for dynamic delegation and identity creation in public key infrastructures - highly customizable, open source, and easy to install - the package setup requires a few MB.
The main differences with respect to a native User Interface - the EGEE/EGI Grid access point - are the extreme ease of use and the no-need of users registration. This way the end user needs only his/her personal X.509 certificate, issued from a Certification Authority, and an access to the Internet. The X.509 personal certificate does not get out from the local machine, strictly compliant to the EGEE/EGI policies, and the gLite [5] User Interface commands are split into client and server, increasing the security level.

An extra security improvement has been achieved by implementing a client - server mechanism for proxy dynamic delegation, avoiding the need of connection to the MyProxy server for usual job operation. It allows to reduce the time needed for the job submission, a higher efficiency and an improved security level in proxy delegation.

## 2. L-GRID architecture and features

The system, implemented as client-server architecture, is based on the Globus gLite Grid middleware and Java Commodity Grid Kits (CoG) library. The client side application is based on a java applet, running both on Windows, Linux and Mac operating systems; it only needs a Web browser connected to the Internet. The server relies on a Globus - gLite User Interface with a Web portal provided by an Apache/Tomcat server.

The novel idea introduced by L-GRID [6] has been the split into client and server of the main *gLite User Interface* commands. This entails an increased security level in Proxy Certificates management: it is no more necessary to send username and password to the MyProxy server for delegating a proxy certificate or have a copy of the X.509 personal certificate on the User Interface. At the same time a smallest amount of data (applications and certificates) needs to to be transmitted over the network: the job input and output files are automatically compressed.

In L-GRID user ID is derived from the Distinguished Name (DN) of the X.509 personal certificate. The user ID is necessary for the server in order to identify the and map the user to a private virtual home directory, distinct it from the others. Within Grid two certificates may not have the same DN, this will ensure no conflict between users.

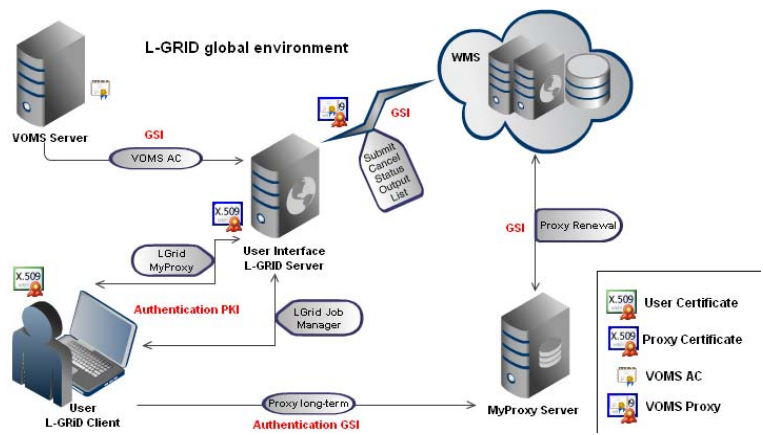

**Figure 1.** L-GRID Architecture.

*2.1. L-GRID Client: a new component in the Grid Architecture*

A completely new component has been introduced in the classical Grid Architecture. While the jobs were usually submitted by a User Interface hosted in a department computing center, the L-GRID solution envisages the use of a light Web client in order to submit the jobs, through a classical User Interface, to the Grid cloud.

The L-GRID client, running on a Web browser and developed as a signed Java applet, downloaded via HTTP protocol or launched using Java Web Start Technology, may be used on whatever Operating System. The Java applet includes a very minimal set of the Globus toolkit libraries, necessaries to implement a secure and trusted message communication between the Web client and the User Interface remotely hosted.

The features implemented on the Web client side are:

- X.509 personal certificate export from .p12 to .pem format - as required by the Globus gLite middleware security system (Grid Security Infrastructure);
- Certificate management and credentials delegation, Automatic Proxy renew;
- Authentication method based on PKI;
- A simple Job Description Language (JDL) editor, in order to ease the end – user in order to write his personal Job;
- Complete lifecycle Job management: Job submission, Status check, Output retrieval.

The last L-GRID release extends our previous work by supporting the management of complex jobs, such as parametric jobs, or jobs collections.

*2.2. L-GRID Server*
The server of the L-GRID portal is embedded in a classical Globus - gLite User Interface. This has been achieved by introducing a simple wrapper to the standard Grid commands on the server side, running L-GRID MyProxy service and L-Grid JobManager service.

As described below, we implemented a fully working mechanism for dynamic delegation of proxy certificate over network. This is done without the exchange of the private key, in order to enhance the security level and reduce the time required for a job submission, compared to an external third party MyProxy server utilization.

The reasons we decided to implement such a mechanism and not to rely on the gLite MyProxy service are primarily dictated by:
- Efficiency: the communications overhead in authentication and proxy certificate recovery is reduced by almost 20%;
- Accessibility: the L-GRID MyProxy service is listening on a standard port, while MyProxy server needs the port 7513 to be open;
- Security: the removal of the user credentials delegation to the MyProxy server strongly reduces the risk of user credentials misappropriation.

The L-Grid JobManager service, running as Web server, takes care of the job management and the Virtual Organization Management System (VOMS) authorization.; it performs the request and returns the result to the client.

*2.3. Portal features*
The system is user-friendly, secure - it uses SSL protocol, mechanism for dynamic delegation and identity creation in public key infrastructures - highly customizable, open source, and easy to install; the installation requires a few MB. The X.509 personal certificate does not get out from the local machine, strictly compliant to the EGEE/EGI policies.

**3. Certificate management and security**
The certificate management in the L-GRID portal is split between client and server. Through a PKI secure and trusted communication [7], the client performs a dynamic credentials delegation toward the server or a third party Grid component - the MyProxy server [8].

*3.1. Credentials Delegation*
The Figure 2 shows the steps involved in the delegation of privileges by creation of a Proxy Certificate between L-GRID Client and L-GRID MyProxy service included in the server side portal.
Below a synthetic description of the steps involved in the Proxy Certificate file generation, as recommended by Ian Foster, Carl Kesselman et al. in *"X.509 Proxy Certificates for dynamic delegation"* [9]:

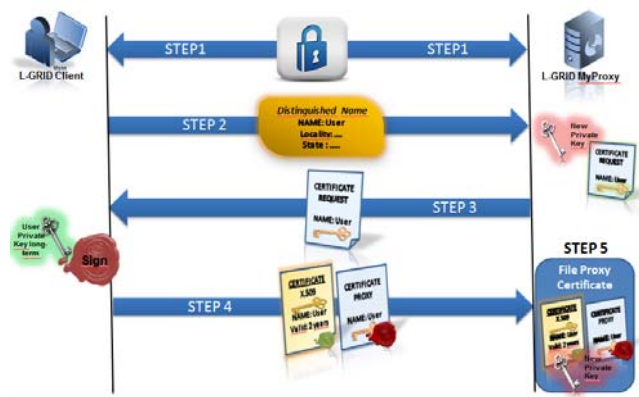

**Figure 2.** Delegation of a Proxy Certificate between Client and MyProxy service.

1. An SSL communication channel is established to perform the required mutual authentication between L-GRID Client and L-GRID MyProxy service. After authentication, an integrity protected channel is established;
2. The Client sends the Distinguished Name (DN) contained in the User Certificate to the MyProxy service;
3. MyProxy service generates a new public and private key pair. With the new public key, a Certificate Signing Request (CSR) is created - the service uses user DN to fill in appropriate values for the fields in CSR - and sent back over the secured channel to the initiator;
4. The client uses the private key associated with its own User Certificate to sign the certificate request, generating a new Proxy Certificate containing the newly generated public key from MyProxy service;
5. The new Proxy Certificate is sent back over the secured channel to MyProxy service, which places it into a file with the newly generated private key and user public key certificate. This new Proxy Certificate file is then available for use on the target service for applications running on the user's behalf.

## 4. Portal use and performances
The portal is intended to be a helpful tool to access Grid resources shared all around the world via a simple Web interface, using whatever operating system and browser. The end user needs only her/his own X.509 personal certificate, issued from a Certification Authority, while no user registration is required. L-GRID provides the typical operations involved in a Grid environment: certificate conversion, job submission, job status monitoring, and output retrieval. It provides also an easy to use Job Description Language (JDL) editor.

*4.1. L-GRID prototype*
The first running prototype is hosted at the moment at the High Performance Computing Center of the Scuola Normale Superiore HPC@SNS, Pisa, Italy. It is reachable and tested at the address https://gridui.sns.it after trusting the downloaded Java applet.
The full code, completely open source (Apache 2.0 licensed), may be free downloaded at page http://sourceforge.net/projects/l-grid/

*4.2. Graphical interface*
The portal presents to the user a simple and easy browsable graphical interface, intuitive to use for a newbie too. The Figures 3 and 4 show respectively the Job submission interface and the Job status monitoring window, where different colors are associated to different job status: running, successfully done, cleared, aborted.

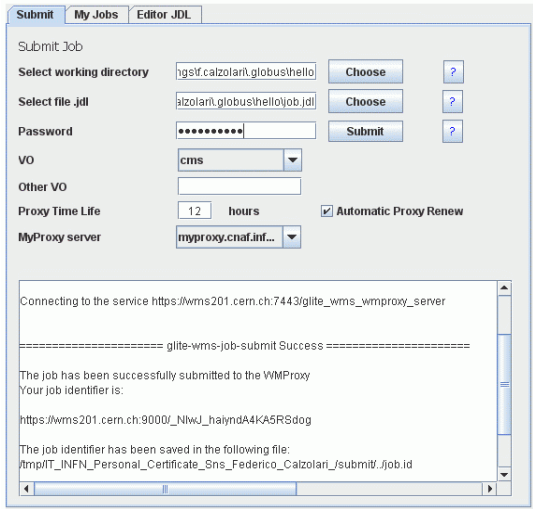 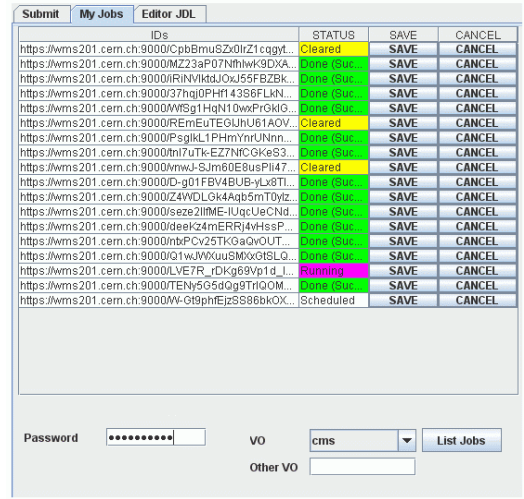

**Figure 3.** Job submission.    **Figure 4.** Job status monitor.

*4.3. Performances*
An extra security improvement has been achieved by implementing a mechanism for dynamic delegation - responsible for the dynamic delegation in proxy certificates - between client and server. In our case, the MyProxy server is contacted only to renew the proxy in long term jobs. It allows to reduce the time spent for the job submission, granting at the same time a higher efficiency and a better security level in proxy delegation and management.
The dynamic delegation implementation into the User Interface + L-GRID server allows to reduce the time spent for the job submission, granting at the same time a higher efficiency.
As shown in the histogram below, the average time for job submission, status monitoring, and output retrieval by using the L-GRID MyProxy service embeded in L-GRD (in blue) is about one second lower than using a classical thirdy part MyProxy server (in red).

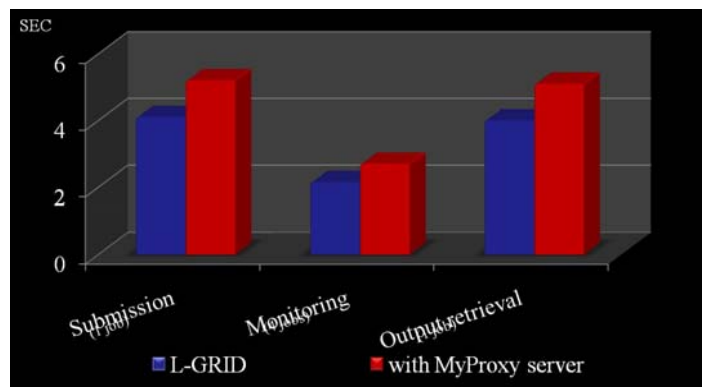

**Figure 5.** Time comparison between L-GRID and classical MyProxy server.

**5. Conclusions and future work**
The L-GRID portal, developed at the Scuola Normale Superiore in collaboration with the University of Pisa (Italy) and the Italian National Institute for Nuclear Research INFN, allows a non expert user to access the Grid resources without any knowledge required on the complex Grid environment. The user is not required to be registered; he only needs his personal X.509 certificate in order to submit a job to the Grid.
By the implementation of mechanism for dynamic delegation directly into the server side portal, the system benefits of an extra improvement in terms of security and a better performance in the time spent for the job management. The portal is fully open source written, light and simple to install and use; it is also fully compliant to the Certification Authority policies with respect to the certificates management.

*5.1. Future developments*
The results obtained encourage future developments. Further steps are represented by the integration with a LDAP Kerberos AAI Authentication Authorization Infrastructure, and the customization for LHC and Theophys Virtual Organizations.